# Effective Data Stewardship in Higher Education: Skills, Competences, and the Emerging Role of Open Data Stewards.

Panos Fitsilis[a*], Vyron Damasiotis[b], Charalampos Dervenis[a],

Vasileios Kyriatzis[c], Paraskevi Tsoutsa[b]

[a]*Business Administration Department, University of Thessaly, Greece;* [b]*Department of Accounting and Finance, University of Thessaly, Greece;* [c]*Department of Digital Systems, University of Thessaly, Greece*
[*] *fitsilis@uth.gr*

**Abstract**

The significance of open data in higher education stems from the changing tendencies towards open science, and open research in higher education encourages new ways of making scientific inquiry more transparent, collaborative and accessible. This study focuses on the critical role of open data stewards in this transition, essential for managing and disseminating research data effectively in universities, while it also highlights the increasing demand for structured training and professional policies for data stewards in academic settings. Building upon this context, the paper investigates the essential skills and competences required for effective data stewardship in higher education institutions by elaborating on a critical literature review, coupled with practical engagement in open data stewardship at universities, provided insights into the roles and responsibilities of data stewards. This approach bridges the theoretical and practical aspects of data stewardship, offering a holistic view of the requirements for effective management and dissemination of data. In response to these identified needs, the paper proposes a structured training framework and comprehensive curriculum for data stewardship, a direct response to the gaps identified in the literature and the practical insights gained from the study. It addresses five key competence categories for open data stewards, aligning them with current trends and essential skills and knowledge in the field, and elaborates on five critical streams of knowledge required to build a comprehensive understanding for open data managers. By advocating for a structured approach to data stewardship education, this work sets the foundation for improved data management in universities and serves as a critical step towards professionalizing the role of data stewards in higher education. The emphasis on the role of open data stewards is expected to advance data accessibility and sharing practices, fostering increased transparency, collaboration, and innovation in academic research. This approach contributes to the evolution of universities into open ecosystems, where there is free flow of data for global education and research advancement.

**Introduction**

The rise of open science represents a profound and impactful change in how research is conducted and disseminated. Based on the fundamental principles of transparency, collaboration, and accessibility, open science seeks to break down the barriers of traditional research, facilitating the unrestricted dissemination of knowledge within both the scientific community and the general public. Early it was identified that this change is reshaping the landscape of academic research, setting new standards for how scientific inquiry and discoveries are shared (Fecher & Friesike, 2014).

A central feature of the open science movement is the concept of open data (Open Knowledge Foundation, 2017). This involves providing free research data that others can analyse, replicate, and disseminate. Open data increases the reproducibility of research findings, fosters innovation through collaborative efforts, ensures efficient use of resources avoiding duplication of effort and necessitates effective implementation of open data practices to realize the full potential of open science (Kassen, 2018).

Universities stand at the forefront of open data production, spanning their efforts across various facets of education, research, and institutional management. Data plays a crucial part in their multifaceted function, encompassing research discoveries, evaluations of quality, specific operational information, learning analytics (West & Paton, 2018), and Open Educational Resources (OERs). Moreover, Universities play a crucial role in the open data ecosystem by serving as centers of knowledge, research, and innovation since they provide valuable contributions to both academic and societal progress (University of Cambridge, 2017; University of California, 2023; Rodriguez, et al. 2020).

In addition, most Higher Education Institutions' efforts, projects and initiatives on open data primarily focus on its provision, encompassing aspects such as data portals and various datasets, rather than on the need for development of human capital in form of the necessary knowledge and skills. This disparity underscores a domain in which colleges may assume a pivotal and important function. Universities must urgently include "open data" concepts into their curricula to promote the development of "open data" human capital. This initiative will not only improve comprehension and proficiency in this domain, but also fulfill the increasing need for open data knowledge among students and professionals (Papageorgiou et al. 2023).

The open data produced by universities in this age of information serves as evidence of their dedication to openness and cooperative advancement (Ramírez & Tejada, 2018). Research data has the potential to result in novel scientific findings (Shu & Ye, 2023), while the evaluation of data quality guarantees high educational standards (Pampel & Dallmeier-Tiessen, 2014). Operational data, on the other hand, facilitates effective institutional administration, whereas learning data enables tailored educational experiences. In addition, the utilization of Open Educational Resources (OERs) promotes equal access to education, as highlighted by (Hansen and Reich, 2015). This further solidifies the role of universities in building a connected and knowledgeable global community. The implementation of an open data approach represents the conversion of institutions into open ecosystems, distinguished by the uninhibited exchange of information, hence promoting innovation and societal advancement. Universities are vital in an open ecosystem as they not only provide knowledge but also actively facilitate cooperation and community engagement. This enables them to advance the boundaries of global education and research (Zubcodd et al, 2016).

Open data stewards play a crucial role in facilitating the management, archiving, accessibility, quality, integrity, and ethical compliance of data. They serve as intermediaries between those who generate data and those who utilize it, assuming a vital responsibility in the management of open data (Rosenbaum, 2010). Pascu and Burgelman (2022) highlight the significance of data stewards in facilitating the shift towards open data. They contend that a significant obstacle to implementing open data is the time needed to acquire the requisite skills and experience. They predict that Europe alone needs around 500,000 data stewards, emphasizing the magnitude of the demand. The importance of investing in data stewardship skills to enhance the speed and efficiency of scientific research is emphasized by this assessment, suggesting that dedicating just a small percentage of the funds given for data infrastructures might provide considerable assistance for this essential role in open science (Mons, 2018).

This study aims to investigate the essential skills and competences required by open data stewards in universities and research educational organizations to effectively support the open data aspect of open science. We investigate the essential skills and expertise required for open data stewards to efficiently oversee, curate, and facilitate the utilization of open data in academic and research environments. The study commenced with a narrative literature review, which is well-suited for complex and interdisciplinary domains such as open data stewardship.

This approach facilitated a thorough comprehension of the subject matter from various viewpoints. The process progressed by actively participating in open data stewardship at universities, gaining valuable understanding of the duties and obligations of data stewards, followed by conducting scholarly research on the latest trends in the field. Through careful analysis, five crucial competence categories for open data stewards were identified. These categories were determined based on their relevance to responsibilities, coverage of data management aspects, alignment with current trends, and emphasis on essential skills and knowledge. Finally, the study encompassed the creation of a curriculum for open data stewardship, which emphasized on the identification of the intended audience, the establishment of learning goals, and the design of an interactive and captivating learning environment.

This paper aims to fill a notable gap found in the existing literature: the absence of structured training programs and comprehensive curricula tailored for data stewardship, an emerging discipline within higher education. Although data stewards have been acknowledged as crucial (Borgman, 2018), there is a perceptible lack of a clearly defined occupational profile or job description for data stewards in university settings. This paper presents a comprehensive framework for training and developing data stewards in higher education. This document highlights the necessary skills and competences required for this role, with a specific emphasis on the dynamic field of open data stewardship.

The paper is structured as follows: the introduction, outlining the context and importance of open science and open data; Section 2 is the background, providing a comprehensive overview of open data and its implications in research; Section 3 is the research methodology, detailing the approach and methods used in our analysis; Section 4 named Findings and Discussion, is presenting the core competences and skills identified for open data stewards for HEIs; section 5 presents a sample curriculum design for open data stewards with respective modules; sections 6 presents the conclusions, summarizing the insights gained and finally section 7 present research limitations and implications for the future of open data in universities and in science in general.

**Background**

As more universities produce diverse sets of data, the role of higher education institutions in generating wide-ranging sets of information is becoming increasingly central. But it is not just a product of research; it serves to stimulate further knowledge exploration. FAIR (Findable, Accessible, Interoperable, and Reusable) principles emphasize the importance of making data reusable and adaptable for future research initiatives (Wilkinson, 2016). Detailed documentation guides researchers on how to extract value from the data, while open licenses ensure its accessibility and unlimited use while this commitment to reuse fosters a virtuous cycle of scientific discovery. By adhering to FAIR principles, researchers are able to enhance the influence of their work. This is achieved by ensuring that their data is easily discoverable, accessible, interoperable, and reusable. Consequently, the visibility of their data is heightened, its reach is expanded, and the potential for groundbreaking discoveries is facilitated. In this data-driven world, FAIR principles are the compass that guides researchers towards a future of seamless collaboration, innovation, and scientific advancement (Mons, 2018; European Commission, 2019).

The data generated by HEIs possess a multifaceted and diverse character that extends beyond the traditional confines of academic research outputs. This encompasses a diverse array of data pertaining to educational procedures, research endeavors, and administrative operations. Furthermore, the data from HEIs is not only crucial for making internal academic and operational decisions, but it also has significant importance for external stakeholders such as policymakers, fellow researchers, and the general public. This data enables informed decision-making and contributes to societal progress. Nevertheless, the process of achieving efficient open data generation and utilization in HEIs is a challenging one, with obstacles to overcome such as technical barriers, apprehensions regarding data accuracy and confidentiality, absence of standardized protocols, and a scarcity of adequately skilled personnel to proficiently oversee and curate this data. (Rodriguez,2020; Zhu, 2020).

In essence, HEIs play a crucial role in generating diverse data that contribute to their multifaceted functions in teaching, research, and social progress. The data generated by these institutions can be categorized into four main groups: (i) Open Research Data, which consists of data from scientific studies and research projects; (ii) Open Educational Data, which includes resources used in teaching and learning; (iii) Open Operational Data, which is related to the administrative and operational aspects of the institutions; and (iv) Quality Assurance Data, which pertains to the evaluation and enhancement of institutional practices and standards. Each of these categories is further expounded upon in the following sections.

*Open Research Data*

According to the UNESCO Recommendation (UNESCO), open science is defined as a set of principles and practices designed to make scientific research from all fields accessible to everyone. This approach benefits both scientists and society. UNESCO sees open science as a means to reduce inequalities both within and between countries while it aligns with the Universal Declaration on Human Rights' Article 27, which asserts that "everyone has the right to freely participate in community cultural life, enjoy the arts, and share in scientific advancements and its benefits."

Moreover, one of the four pillars of open science is the dissemination of open scientific knowledge, wherein open research data is identified as a key constituent. This emphasizes the critical role of freely accessible research data in the broader framework of open science, underscoring its importance in fostering a more inclusive and collaborative scientific community (European Commission, 2023). Research data consists of data obtained from diverse scholarly research endeavors. The data encompasses several types, such as experimental, observational, simulation, and derived or compiled data. The availability of research data is crucial for scientific breakthroughs and advancements, and the act of openly sharing this data helps expedite the pace of research.

Therefore, the availability of open access research data is vital, as Borgman (2012) emphasizes the importance of open access to research data by stating that it is:

- **Essential** to scientific progress. When researchers can freely share their data, it allows others to build upon their work and make new discoveries. This can lead to faster and more significant advancements in science.
- More **efficient**. When researchers don't have to spend time and resources duplicating data that is already available, they can focus on new research questions. This can lead to a more productive scientific community.
- More **equitable**. When all researchers have access to the same data, it levels the playing field and allows researchers from all backgrounds to contribute to scientific progress. This can help to ensure that scientific discoveries are not limited to those who have the resources to access proprietary data.
- More **transparent**. When data is publicly available, it is easier for other researchers to check the validity of the findings. This can help to ensure that scientific research is accurate and reliable.
- More **compliant** with ethical guidelines. Many ethical guidelines require that researchers make their data available to others. Open access to research data can help to ensure that researchers are complying with these guidelines.

Open research data efforts, such as the European Open Science Cloud (EOSC), seek to enhance the accessibility and reusability of research data for academics, policymakers, and the public. The EOSC is a federated infrastructure that offers researchers a centralized platform for conveniently accessing, organizing, and collaborating on their research data, bringing together a network of data repositories, services, and tools from many locations across Europe, having as an ultimate objective to facilitate the discovery and utilization of open research data by researchers, as well as to foster the creation of novel tools and services for managing open research data. This ongoing initiative, although being ambitious and in the developmental stage, holds the capability to revolutionize the management and sharing of research data throughout Europe. By providing a centralized platform for open research data, the EOSC can help to accelerate scientific progress, foster collaboration, and improve the quality of research (European Commission, 2021; European Commission, 2016).

In addition to the EOSC, there are a number of other key initiatives for open research data (Hansson & Dahlgren, 2022), including:

- The Dataverse Network: A digital commons of social science research data that provides a centralized space for researchers to share and find data (https://dataverse.org/).
- Zenodo: An open-access repository for scientific data and publications (https://zenodo.org/).
- Pangaea: A repository for geoscientific data (Felden, et al., 2023).
- The International Data Rescue Initiative (IDRI): A non-profit organization that collects and preserves open access data. It provides a single point of access for information on the status of past and present data rescue efforts worldwide, including the best methods and technologies involved in data rescue, as well as metadata for data that require rescue. (https://www.idare-portal.org/).
- The Earth System Grid Federation (ESGF): A distributed data infrastructure for earth science data (https://esgf.llnl.gov/).

These initiatives contribute to enhancing the accessibility and reusability of open research data, exerting thus a substantial influence on scientific advancement and collaboration.

As the trend of open research data evolves, a range of challenges becomes apparent, for this domain. An important problem is the data gap, which refers to the disparity between the extensive quantity of data being gathered and the relatively smaller fraction that is being put to productive use. This gap underscores a crucial aspect of emphasis in the use of open research data as it is accountable for introducing mistakes, biases, and constraining the extent of insights obtained. An additional factor that significantly influences the situation is the infrastructure, which frequently lacks robustness, thus obstructing the accessibility and usage of data. Equally critical is the lack of standardization in open data, which consists of a variety of formats and protocols that perplex researchers in their pursuit of data coherence. The legal and ethical dilemmas become intricately intertwined in this complex situation, with concerns related to privacy, intellectual property, and the possibility of misuse. Compounding these challenges is a lack of incentives for researchers to venture into open data sharing, shadowed by fears of losing data control (Zuiderwijk, Shinde & Jeng, 2020). Each of these variables intimately contributes to the complicated and promising field of open data, requiring attention and innovation in order to fully achieve its actual potential (Reichman, Jones, & Schildhauer, 2011).

*Open Educational Data*

In the extensive ecosystem of HEIs, educational data plays a crucial function, becoming a vital and inseparable component of academic life, fundamental in influencing the overall educational experience. It includes, but is not limited to, detailed records of course offerings, insightful analytics on learning outcomes, and evaluations from academic assessments. Each of these data points has a role in the functioning of educational processes, leading to a comprehensive knowledge of the educational environment.

Moreover, the emergence of Massive Open Online Courses (MOOCs), micro-learning, and personalized learning has significantly increased the volume and variety of educational data generated by HEIs. MOOCs, with their global reach, contribute extensive data on diverse learning patterns and preferences (Silveira, 2016). Micro-learning, focusing on concise, targeted content, adds to this dataset with insights into effective content delivery and learner engagement (Leo et al., 2020). The tendency of personalized learning, particularly highlighted in the work of (Iatrellis, 2017; Fitsilis, 2022), is a testament of how educational data can be leveraged to tailor learning experiences to individual needs. This approach not only improves the learner's experience but also provides rich, detailed data on individual learning paths, preferences, and outcomes. This rising quantity of data from these innovative educational models is transforming how educational efficacy is understood and improved in HEIs.

In addition, the field of educational data is intrinsically linked to the concept of Open Educational Resources. According to UNESCO (2023) "Open Educational Resources (OER) are learning, teaching and research materials in any format and medium that reside in the public domain or are under copyright that have been released under an open license, that permit no-cost access, re-use, re-purpose, adaptation and redistribution by others". OERs represent a revolutionary approach in education, characterized by freely accessible, openly licensed text, media, and other digital assets that are useful for teaching, learning, and assessing as well as for research purposes. The proliferation of OERs has been a game-changer, democratizing access to educational resources and fostering an environment of inclusivity and collaboration (Atkins, Brown, & Hammond, 2007).

HEIs are increasingly managing repositories of OERs, making them a treasure trove for educators and learners alike. These repositories serve as hubs where educational materials can be shared, discovered, and utilized to enrich the learning process. They range from complete course materials to modules, textbooks, streaming videos, tests, software, and any other tools, materials, or techniques used to support access to knowledge. OERs are high-quality, freely accessible educational resources that can be used for teaching, learning, or research and they are organized as online platforms with repositories that aggregate and organize OER from various sources. The following list is an indicative list of OER's repositories:

- MERLOT (Multimedia Educational Resource for Learning and Online Teaching) is a comprehensive collection of high-quality, peer-reviewed educational resources, including full courses, modules, tutorials, simulations, and multimedia content. These resources cover a wide range of subjects and are designed for use by learners of all ages and backgrounds. ( https://merlot.org/merlot/)

- The OER Commons is a collaborative project of the Hewlett Foundation and the William and Flora Hewlett Foundation to develop and promote the use of high-quality, open educational resources. The repository contains a wide variety of resources, including textbooks, course materials, and teaching tools. (https://oercommons.org/)

- The OpenCourseWare Consortium is a global community of higher education institutions that have committed to making their course materials openly available to everyone. The consortium's website contains a vast collection of open course materials, covering a wide range of subjects and levels. (https://ocw.mit.edu/)

- Curriki is a non-profit organization that develops and provides high-quality, open educational resources for pre-K-12 education. The organization's website contains a variety of resources, including lesson plans, activities, and assessments. (https://www.curriki.org/)
- Khan Academy is a non-profit organization that provides free educational resources for learners of all ages. The organization's website contains a vast library of videos, exercises, and practice problems covering a wide range of subjects. (https://el.khanacademy.org/)

Essentially, when educational data intertwines with the expanding trend of OERs, it serves as the foundational pillar of contemporary education systems within HEIs. This symbiotic relationship showcases the transformative power of data in reshaping educational landscapes for more informed, efficient, and inclusive educational methodologies.

However, open education face challenges that can hinder their growth. For example, the lack of awareness about copyright issues by academics is serious challenge especially in the domain of digital content as many academics are unfamiliar with how to retain certain rights while sharing their work. Open licenses such as Creative Commons address this need but require authors to relinquish some control and navigate complex rights issues. Another challenge is the sustainability of all these repositories as their development during the last years has created a great competition for funding. Finally, ensuring OER's quality is another challenge have to face students, teachers and self-learners. (Hylén, n.d.).

*Open* **Operational** *Data*

The utilization of accessible operational data is crucial at HEIs since it facilitates transparency, improves decision-making, aids research and innovation, encourages civic participation, and stimulates economic growth. According to Janssen et al. (2012), open data analysis acts as a catalyst for improving the effectiveness, accountability, and creativity of institutions, thus promoting a more open and adaptive learning environment. By making institutional data accessible to the public, HEIs can improve the performance, accountability, and overall impact of higher education. (Iatrellis, 2019; Iatrellis, 2020) provide a holistic view of the role of open operational data in higher education, emphasizing the potential of this data to enhance student success and institutional effectiveness. They propose a framework that utilizes machine learning and semantics to analyze and interpret this data, enabling universities to make informed decisions about curriculum design, teaching methods, and resource allocation.

Open operational data, encompassing a multitude of operational facets, is indispensable for the efficacious governance and administration of HEIs. Comprising elements such as student records, staff information, financial data, and facilities management records, is crucial for guiding HEIs towards operational excellence and institutional efficacy. Exploring the domain of student records, one discovers a vast repository of data illuminating student demographics, academic pathways, and engagement patterns. This data is more than just administrative tools, serving as pivotal instruments in monitoring and predicting academic progression, gauging student needs, and strategizing interventions aimed at amplifying student success and retention. In parallel, the corpus of staff information, encompassing data on faculty and administrative personnel, emerges as a cornerstone in the domain of human resource management. This data empowers informed staffing decisions, fosters an environment conducive to professional growth, and aligns human resources with the overarching academic objectives of the institution.

The financial data layer, defining budgetary allocations, spending trends, and revenue channels, forms the financial backbone of HEIs. It plays an important role in formulating budgetary strategies, ensuring judicious allocation of resources, and sustaining the financial health of the institution. The integration of this financial data with the strategic objectives of the HEIs facilitates well-informed decision-making, which is crucial for ensuring the financial stability and growth of these educational institutions. Some examples of online services offering such data are:

- IPEDS (Integrated Postsecondary Education Data System) which is a database of information on colleges and universities in the United States. It includes data on student enrollment, finances, and other areas. IPEDS data is available from the National Center for Education Statistics (NCES). (https://nces.ed.gov/ipeds/)

- CollegeScorecard.org: This is a website from the U.S. Department of Education that provides information on colleges and universities. It includes data on student outcomes, finances, and other areas. (CollegeScorecard.org)

- Centre for Institutions and Governance (CIG) was founded to address the need for high-quality, credible research and education on corporate governance issues in Hong Kong, China and Asia. (https://www.bschool.cuhk.edu.hk/centres/centre-for-institutions-and-governance/)

Finally, data related to facilities management encapsulates an array of information crucial for the upkeep, optimization, and expansion of institutional infrastructure. This data, ranging from maintenance records to space utilization metrics, is pivotal in ensuring that the physical infrastructure of HEIs is aligned with their educational and research missions. It facilitates the creation of an environment that not only supports but also enhances the academic and research pursuits of the institution.

Managing open operational data in higher education is a task with critical and complex challenges. Beyond the obvious need to ensure privacy and security for sensitive information of students and staff the need to maintaining the accuracy and relevance of this data is also a significant task, demanding advanced analytical skills which may not be readily available. Another noteworthy challenge is that institutions frequently prioritize compliance above strategic decision-making, which restricts the time available for deriving data-driven insights. Furthermore, analytics teams frequently operate in silos within specific departments, hindering the widespread adoption of a data-driven culture across the institution. Finally, integrating legacy data systems presents technical difficulties, delaying the realization of the benefits of analytics. (Krawitz et al., n.d.)

In summary, open administrative data in HEIs serves as more than just a storage facility for data. It is a dynamic and strategic resource, playing a crucial role in connecting and coordinating the several operational aspects of HEIs, guaranteeing their smooth functioning, strategic expansion, and alignment with their primary academic and research goals.

*Quality Assurance Data*

Quality-related open data produced in universities represents a critical aspect of the open data landscape within HEIs. This data contains a wide range of information that directly relates to the quality of education, research outcomes, institutional performance, and compliance with accreditation standards.

- Academic Quality Data: This includes data on program effectiveness, course completion rates, student satisfaction surveys, and alumni outcomes. Such data is vital in assessing and improving the quality of educational programs offered by universities. It provides insights into the effectiveness of teaching methodologies, curriculum relevance, and overall student engagement and satisfaction. Examples of such data include Graduation rates, degree completion rates, average debt levels, etc.

- Research Quality Data: In this category, data encompasses research outputs, publication metrics, citation analyses, and research impact. This data helps in evaluating the quality and impact of research conducted at universities. Metrics such as the number of

- publications in high-impact journals, citation counts, and research grants and patterns awarded serve as indicators of research excellence and innovation.

- Institutional Performance Data: This includes data related to university rankings, accreditation outcomes, and compliance with educational standards. This data is used by universities to benchmark their performance against national and international standards, ensuring continuous improvement and adherence to high-quality standards in education and research.

- Quality Assurance and Accreditation Data: Data from internal and external audits, accreditation reviews, and quality assurance assessments are pivotal in maintaining and enhancing the educational standards of HEIs. They provide a framework for continuous improvement, helping universities to identify areas of strength and opportunities for development.

Regarding institutional performance data, the renowned Times Higher Education (2023) University Ranking list, considered a global benchmark for academic excellence, predominantly relies on data directly obtained from universities and publishing institutions. This methodology, while comprehensive, does not predominantly utilize open data that is publicly available for cross-verification and scrutiny. The data utilized for these rankings, which includes several factors such as academic repute and research productivity, is predominantly inaccessible for public verification. This practice emphasizes a notable deficiency in the openness of ranking procedures and emphasizes the necessity for such influential data to be readily accessible. Publicly disclosing this data will not only bolster the credibility and responsibility of the ranking system, but also conform to the wider principles of transparency and unrestricted availability in the academic community. The adoption of open data in university rankings has the potential to establish a standard for evaluation processes in higher education that are both transparent and inclusive. Quality-related open data is essential for promoting openness and accountability in institutions. It allows stakeholders, like students, professors, researchers, and policymakers, to make well-informed decisions using concrete metrics for quality and performance. By making this data openly accessible, universities not only demonstrate a commitment to excellence but also contribute to the broader goal of enhancing the quality of higher education globally.

The production and management of these data types present both opportunities and challenges for HEIs. Open data initiatives, such as those adopted by the University of Cambridge (2023) and the University of California, highlight the growing interest in making

HEI-generated data more accessible and usable for a broader audience. However, while open data has the potential to enhance transparency, accountability, and innovation, reaping its full benefits for society and the academic community necessitates a strategic approach to data management, stewardship, and governance.

Among the various challenges faced by open data, the need for data standardization stands out. Data from different higher education institutions may use varying formats, making meaningful comparisons and benchmarking extremely difficult, if not impossible. Clear policies and procedures are essential to manage data, ensure compliance with regulations, and address issues such as data ownership and usage rights, enabling benchmarking and continuous improvement. (Cai and Zhu, 2015; McDonald, 2022).

The role of HEIs in producing and disseminating open data is thus crucial and multifaceted, requiring a strategic approach to overcome existing barriers and fully leverage the potential of open data. Figure 1 highlights the key findings for HEIs as open data producers.

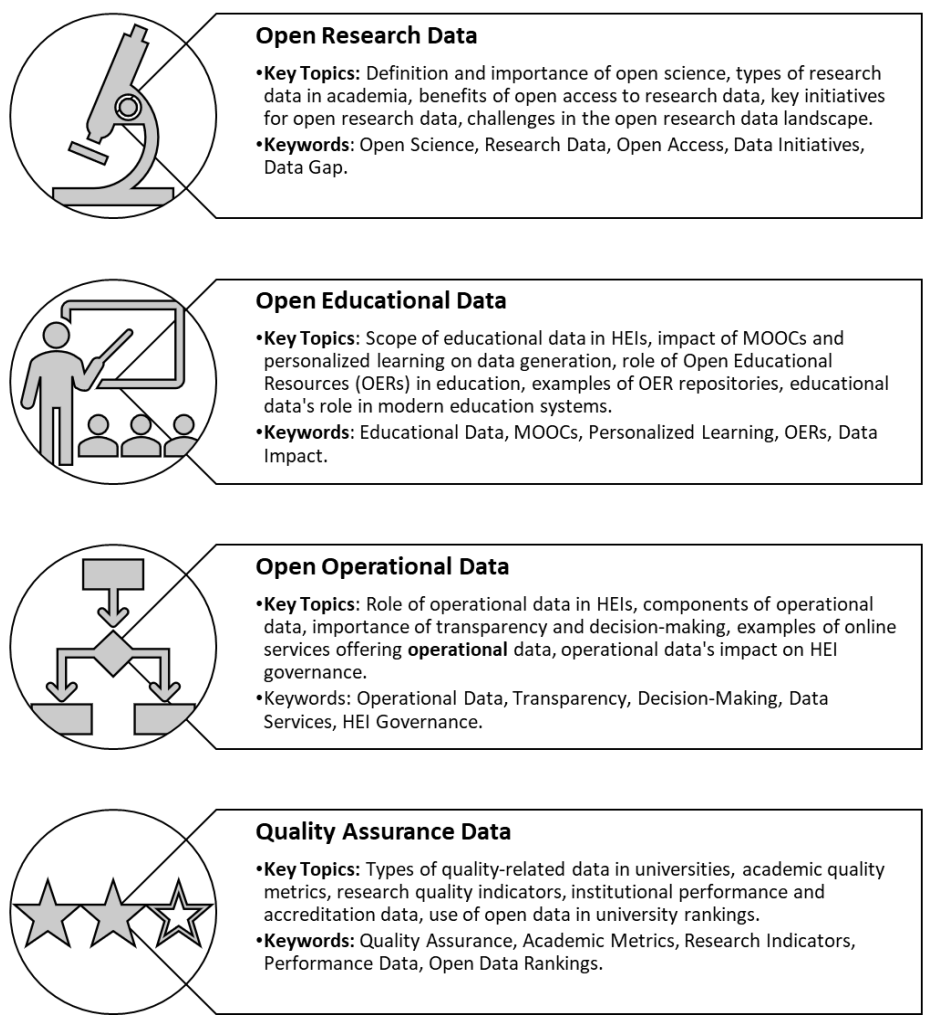

*Figure 1: Higher Education Institutions as Open Data Producers*

**Research methodology**

In conducting this research, we adopted a constructivist approach, viewing research as a product shaped by specific times, social conditions, and interactional situations (Charmaz, 2008). This perspective acknowledges that while researchers' prior experiences and perspectives guide their research, they do not predetermine its course. Thus, our accumulated experience and insights gained from working in the domain of open data stewardship in universities have acted as preliminary drivers in our research.

Our initial step was to conduct a narrative literature review. This type of review, unlike systematic reviews, does not start with a predetermined research question or a specified search strategy but rather with a general topic of interest (Demiris et al., 2019). Such reviews are particularly suited to complex and interdisciplinary fields like open data stewardship, allowing for a comprehensive understanding of the topic from multiple perspectives (Pae, 2015).

In our investigation, we cohered to the principles of Grounded Theory, allowing themes to emerge naturally from the literature. This approach aligns with the constructivist perspective, as it allows for the content and structure of the inquiry to be emergent, based on the data (Charmaz, 2008). We did not set specific hypotheses before conducting the literature review. Instead, we collected and analyzed data simultaneously, looking for themes and patterns that emerged organically. Furthermore, our research process unfolded iteratively across the stages, reflecting the evolving nature of the field of open data stewardship. As new technologies, policies, and challenges continuously reshape this domain, it becomes imperative to adapt and evolve the research focus to stay relevant. According to Charmaz (2006), the inductive and iterative process of Grounded Theory analysis keeps the research focused and incisive, allowing for a higher level of abstraction and theoretical exploration.

**Stage 1: Practical engagement in open data stewardship at universities.**

To concretely begin our research, we drew on our direct involvement in various projects and initiatives related to open data stewardship at universities. These engagements provided practical insights into the roles and responsibilities of open data stewards, shaping our initial understanding and guiding our focus on this emerging professional profile. This practical foundation, enriched by our hands-on experience, offered a unique lens through which to view and understand the intricacies of data stewardship in academic settings.

**Stage 2: Scholarly Inquiry for exploring current trends in open data stewardship.**

Building upon this practical foundation, the study broadened its scope to include current trends and advancements in open data stewardship, thus enriching the initial understanding with a forward-looking perspective. This phase marked the transition from practical engagement to structured academic exploration, integrating experiential knowledge with scholarly research.

The literature review was conducted through a narrative approach, examining each aspect of open data stewardship through existing academic literature, case studies, and relevant institutional documents. This methodology provided a rich, multi-faceted view, connecting theoretical frameworks with practical implementations in the university context.

This review resulted in five categories of competences for open data stewards, presented in the following section. The key criteria for selecting these categories were:

- Relevance to the core responsibilities of open data stewards: The competences should directly address the essential tasks and challenges faced by open data stewards in managing and promoting open data within universities.

- Coverage of the multifaceted nature of open data management: The categories should encompass the various aspects of open data governance, including technical, legal, domain-specific, analytical, communication, and project management aspects.

- Alignment with current trends and best practices: The competences should reflect the latest developments in open data practices and reflect the best practices adopted by leading institutions.

- Emphasis on skills and knowledge required for effective open data stewardship: The competences should focus on the skills and knowledge that are essential for open data stewards to make informed decisions, collaborate effectively, and contribute to the successful implementation of open data initiatives within universities.

By considering these criteria, five categories of competences for open data stewards were identified as the most comprehensive and relevant framework for developing and equipping individuals with the necessary skills and knowledge to effectively manage and promote open data within universities.

**Stage 3: Developing a curriculum to support open data stewardship.**

Curriculum development is a multidimensional effort, comprising of a sequence of interconnected steps, each contributing to the creation of an effective and engaging learning experience. This process is characterized by an iterative nature, requiring continuous refinement and adaptation to ensure alignment with the evolving needs of the target audience and the ever-changing landscape of open data stewardship.

The first critical step in this process lies in meticulously defining the target audience. In our specific context, the intended participants encompass a diverse range of individuals and institutions, including HEIs workforce, educational organizations, and entities primarily dedicated to fostering open science principles. Understanding the specific knowledge, skills, and experiences of this audience is crucial to tailor the curriculum's content and delivery effectively.

Once the target audience has been identified, the next step was to articulate clear and measurable learning objectives. These objectives serve as the guiding framework for the curriculum's development, ensuring that it is focused on equipping participants with the competences essential for effective open data stewardship.

The curriculum should be a flexible and captivating collection of learning activities, integrating various teaching methods to accommodate different learning styles and preferences. Lectures, workshops, case studies, and hands-on exercises should interweave seamlessly, fostering active participation, critical thinking, and practical application of knowledge. Additionally, assessment methods should be carefully chosen to align with the learning objectives and effectively gauge the participants' mastery of the subject matter. A combination of written exams, case studies, and presentations can provide comprehensive feedback and enable continuous improvement of the curriculum.

While this study primarily focuses on the development of the curriculum, it is essential to acknowledge the iterative nature of the process. Regular review and piloting of the curriculum are paramount to ensure its effectiveness and relevance in the ever-evolving landscape of open data stewardship.

**Analysis of the literature review**

Open data stewards are critical in managing and promoting the responsible use of open data within universities. They need to possess a comprehensive set of skills and knowledge to effectively handle the complexities of open data governance, dissemination, and utilization. The five categories of competences emerged represent the crucial areas where open data stewards should direct their expertise as follows:

- **Data Technical Competences:** Open data stewards must have a solid understanding of data management and storage principles, data cleansing and preprocessing techniques, programming skills for data analysis and mining, and the ability to integrate

- **Legal and Ethical Competences:** Open data stewards must be well-versed in data protection and privacy laws, intellectual property regulations, and ethical considerations related to the responsible use of open data. They need to understand the legal and ethical implications of sharing data with the public and ensuring that data privacy is protected while maximizing the benefits of open data.

- **University Domain-Specific Competences:** Open data stewards should have a deep understanding of the specific needs and requirements of the research domains or fields of study within the university. This includes knowing how to manage research data, supporting educational development, structuring learning management systems, and adhering to the principles of open science.

- **Data Analysis and Interpretation Competences:** Open data stewards should be proficient in data analysis techniques, data visualization methods, and data interpretation skills. They should be able to extract meaningful insights from open data and communicate these findings effectively to various stakeholders.

- **Communication, Collaboration, and Project Management Competences:** Open data stewards need strong communication, collaboration, and project management skills to effectively disseminate and promote open data. They should be able to work with researchers, stakeholders, and communities to share data, knowledge, and insights. Additionally, they should have the ability to lead data initiatives and manage projects from conception to implementation, including budgeting, timeline management, and stakeholder engagement.

In summary, the five categories of competences for open data stewards provide a comprehensive framework for developing the skills and knowledge necessary to effectively manage and promote the responsible use of open data within universities. These competences address the technical, legal, domain-specific, analytical, communication, and project management aspects of open data governance. By acquiring and applying these competences, open data stewards can make significant contributions to the advancement of open science and the creation of a more transparent and knowledge-driven university environment.

heterogeneous repositories and utilize linked data standards. These technical skills are essential for ensuring the quality, accessibility, and interoperability of open data assets.

*Table 1: Data Technical Competencies*

**A. Data Technical Competences**

The ability to collect, process and publish data requires technical skills, such as proficiency in data management and storage, data cleaning and preprocessing and data security.

| Code | Competence | Characteristics | References |
|------|-----------|-----------------|------------|
| A1 | Data management processes | The collection and management of the organization's data, in this context, also encompass concerns regarding data governance and release processes | (Donner, 2023; Hendriyati et al., 2022; Gillenson, 2023; Bunkar & Bhatt, 2020; Tran & Scholtes, 2015; Ismael et al., 2018; Universities UK et al., 2015; Rodriguez-F et al., 2020; Tzitzikas et al., 2021; Radchenko et al., 2018; Zubcoff et al., 2016; Zubcoff et al., 2016) |
| A2 | Data storage | Concerns matters of data storage but also the identification and management of the appropriate infrastructure that will be made available to the organization for long-term storage. | (Rodriguez-F et al., 2020; Bunkar & Bhatt, 2020; Tzitzikas et al., 2021; Radchenko et al., 2018; Zubcoff et al., 2016, Zubcoff et al., 2016) |
| A3 | Data cleansing and preprocessing | Using clean data is an important part of drawing the right conclusions. Errors often could be the result of human mistakes in data entry, such as mistyping or incorrect abbreviations. | (Maharana et al., 2022; Ridzuan & Wan Zainon, 2019; Universities UK et al., 2015; Hesteren & Knippenberg, 2021) |
| A4 | Programming | Open data development and data mining include programming skills and knowledge of programming techniques. | (Hendriyati et al., 2022; Radchenko et al., 2018; Hesteren & Knippenberg, 2021) |
| A5 | Enriching data | One dataset is rarely enough to gain insight. By combining multiple datasets, it is possible to obtain a more detailed picture. | (Universities UK et al., 2015; Tzitzikas et al., 2021) |
| A6 | Integration of heterogeneous repositories | Integrating heterogeneous repositories and publishing their metadata as linked data. | (Hendriyati et al., 2022; Rodriguez-F, Arcos-Medina, Pástor, Oñate, & Gómez, 2020; Tzitzikas, Pitikakis, Giakoumis, |

**A. Data Technical Competences**

| | | | |
|---|---|---|---|
| | | | Varouha, & Karkanaki, 2021; Hesteren & Knippenberg, 2021; Piedra, Tovar, Colomo-Palacios, Lopez-Vargas, & Alexandra Chicaiza, 2014) |
| A7 | Data security | Protecting data from internal or external corruption and illegal access protects a university from financial loss, stakeholders trust degradation, university reputational erosion, etc. | (Donner, 2022; Bunkar and Bhatt, 2020; Tran & Scholtes, 2015; Ismael et al., 2018; Tzitzikas et al., 2021; Hesteren, Knippenberg, 2021; Zubcoff, Vaquer Gregori, Mazón, Maciá Pérez, Garrigós, Fuster-Guilló, & Cárcel Alcover, 2016) |

*Table 2: Legal and Ethical Competencies*

**B. Legal and Ethical Competences**

Ensuring that data is open while also protecting privacy, intellectual property and other legal rights requires knowledge of data protection and privacy laws as well as ethical considerations related to the responsible use of data.

| Code | Competence | Characteristics | References |
|---|---|---|---|
| B1 | Knowledge of data protection laws | Covers the rights, licenses and data protection regarding what people can do with the data. | (Donner, 2022; Tran & Scholtes, 2015; Ismael, Mohd, & Abd Rahim, 2018; Universities UK, Open Data Institute, corp creators, 2015; Tzitzikas, Pitikakis, Giakoumis, Varouha, & Karkanaki, 2021; Radchenko, Koroleva, & Baranov, 2018; Hesteren & Knippenberg, 2021; Zubcoff, Vaquer Gregori, Mazón, Maciá Pérez, Garrigós, Fuster-Guilló, & Cárcel Alcover, 2016) |
| B2 | Knowledge of data privacy laws | Open data may leak valuable information to third parties, e.g. competitors. In this sense, knowledge of intellectual property law in science projects with | (Donner, 2022; Bunkar and Bhatt, 2020; Tran & Scholtes, 2015; Ismael, Mohd, & Abd Rahim, 2018; Universities UK et al., 2015; Rodriguez-F, Arcos-Medina, Pástor, Oñate, & Gómez, 2020; Karmanovskiy, Mouromtsev, Navrotskiy, |

**B. Legal and Ethical Competences**

| | | | |
|---|---|---|---|
| | | companies and with other research organizations is necessary. | Pavlov, & Radchenko, 2016; Radchenko et al., 2018; Hesteren & Knippenberg, 2021; Perkmann & Schildt, 2015; Vicente-Saez, Gustafsson, & Van den Brande, 2020) |
| B3 | Ethical skills | Ethical and commercial interests can affect whether the data could be made open access. | (Donner, 2022; Bunkar and Bhatt, 2020; Tran & Scholtes, 2015; Coughlan, 2020; Borgerud & Borglund, 2020) |

*Table 3: University Domain specific Competencies*

**C. University Domain specific Competences**

Understanding the specific needs and requirements of the research domain or field of study is important for developing relevant data and ensuring that it is of high quality.

| Code | Competence | Characteristics | References |
|---|---|---|---|
| C1 | Managing research data | Transparency and accessibility to science outputs as well as authorization and participation in research. | (Donner, 2022; Bunkar & Bhatt, 2020; Universities UK et al., 2015; Rodriguez-F, Arcos-Medina, Pástor, Oñate, & Gómez, 2020; Tzitzikas, Pitikakis, Giakoumis, Varouha, & Karkanaki, 2021; Hesteren & Knippenberg, 2021; Zubcoff, Vaquer Gregori, Mazón, Maciá Pérez, Garrigós, Fuster-Guilló, & Cárcel Alcover, 2016, Zubcoff et al., 2016 Perkmann & Schildt, 2015; Vicente-Saez, Gustafsson, & Van den Brande, 2020; Borgerud & Borglund, 2020} |
| C2 | Educational development | Monitoring the educational processes within a HEI. | (Donner, 2022; Rodriguez-F, Arcos-Medina, Pástor, Oñate, & Gómez, 2020; Coughlan, 2020; Zubcoff et al., 2016; Atenas, Havemann, & Priego, 2015) |
| C3 | Structuring learning | A learning management system (LMS) is designed to make things | (Hendriyati et al., 2022; Donner, 2022; Bunkar & Bhatt, 2020; |

**C. University Domain specific Competences**

| Code | Competence | Characteristics | References |
|---|---|---|---|
| | management systems | easier for all stakeholders by using it to plan, deliver and evaluate a learning process. However, it needs appropriate people to structure and manage it so that it functions properly. | Zubcoff et al., 2016; Atenas et al., 2015) |
| C4 | Understanding Open science principles | Concerns key principles of open science that direct the work of research teams at universities such as, transparency, accessibility to science outputs, authorization and participation to science outputs. | (Gallagher et al., 2020; Rodriguez-F, Arcos-Medina, Pástor, Oñate, & Gómez, 2020; Tzitzikas, Pitikakis, Giakoumis, Varouha, & Karkanaki, 2021; Hesteren & Knippenberg, 2021; Karmanovskiy et al., 2016; Zubcoff et al., 2016; Vicente-Saéz et al., 2020; Atenas et al., 2015) |

*Table 4: Data Analysis and Interpretation Competencies*

**D. Data Analysis and Interpretation Competences**

The ability to analyze and interpret data, identify trends and patterns and draw meaningful conclusions is critical for making effective use of open data.

| Code | Competence | Characteristics | References |
|---|---|---|---|
| D1 | Data analysis | Concerns providing statistics through analyzing raw data to discover trends and metrics, make predictions and eventually make conclusions about the hidden information included in useful datasets regarding students, research, finances, etc. | (Nag & Ahmad Malik, 2023; Bunkar & Bhatt, 2020; Tzitzikas et al., 2021; Radchenko et al., 2018; Zubcoff et al., 2016) |
| D2 | Visualizing data | Visualizations are a useful way of interpreting data, and they help us to unlock insight. | (Universities UK et al., 2015; Tzitzikas et al., 2021; Radchenko et al., 2018; Hesteren & Knippenberg, 2021) |
| D3 | Interpreting data | This is an important feature as it allows users to understand and | (Nag & Ahmad Malik, 2023; Universities UK et al., 2015; |

**D. Data Analysis and Interpretation Competences**

| | | interpret data through previews and visualizations. | Hesteren & Knippenberg, 2021) |

*Table 4: Communication, Collaboration and Project Management Competencies*

**E. Communication, Collaboration and Project Management Competences**

Collaborating with other researchers, stakeholders, and communities to share data and knowledge and to communicate the results and insights from open data is essential. This competences category, also refers to the ability to lead data initiatives and manage projects from conception to implementation including managing budgets, timelines, long-term plans, developing, etc.

| Code | Competence | Characteristics | References |
|---|---|---|---|
| E1 | Dissemination and Communication | An important action within the process of opening data is its dissemination and communication, which will give greater visibility and obtain feedback useful in improving services. | (Coughlan, 2020; Radchenko et al., 2018; Zubcoff et al., 2016; Perkmann & Schildt, 2015) |
| E2 | Publication of open data | Concerns tools used in the context of data publishing, linking, promotion, etc. | (Rodriguez-F et al., 2020; Radchenko et al., 2018; Hesteren & Knippenberg, 2021; Zubcoff et al., 2016; Perkmann & Schildt, 2015) |
| E3 | Collaboration | Collaboration is necessitated by the use of open data, as they not only enhance an organization's transparency through the potential for data reuse in future but also fosters a collaborative environment that accelerates innovation. | (Rodriguez-F et al., 2020; Coughlan, 2020; Radchenko et al., 2018; Hesteren & Knippenberg, 2021; Zubcoff et al., 2016; Vicente-Saez et al., 2020) |
| E4 | Standardization | Concerns defining and understanding standards that are technical documents designed to be used as a rule, guideline, specification, definition and publishing | (Rodriguez-F et al., 2020; Hesteren & Knippenberg, 2021; Zubcoff et al., 2016) |

**E. Communication, Collaboration and Project Management Competences**

| | | protocols. They are agreed upon by consensus and can be used consistently to ensure that products, processes and services are fit for their purpose. | |
|---|---|---|---|
| E5 | Leadership and project management | The ability to lead data initiatives and manage projects from conception to implementation. This includes the ability to develop and manage budgets, timelines and to work with different stakeholders. | Radchenko, Koroleva, & Baranov, 2018; Zubcoff et al, 2016; Vicente-Saez, Gustafsson, & Van den Brande, 2020) |
| E6 | Strategic thinking | The ability to think strategically and develop long-term plans for data management and use. | (Hesteren, & Knippenberg, 2021, Zubcoff et al, 2016; Vicente-Saez, Gustafsson, & Van den Brande, 2020) |

**Sample curriculum for open data stewards**

In this section, we introduce a sample curriculum for open data stewards, emphasizing that this is just one of many potential variations. Depending on the audience type, knowledge level, and specific needs, tailored iterations of this curriculum can be developed, ensuring that each version is optimally designed to cater to the unique educational requirements and learning styles of different groups.

The curriculum has been developed as a result of the research methodology and thorough analysis detailed in the previous sections. It reflects the synthesis of relevant works, tailored to meet the diverse needs and knowledge levels of different audiences. This example encompasses five key subsections: Introductory Modules, Data Management and Exploitation, Management Modules, Legal Issues and Ethics, and Higher Education Data Challenges and Case Studies. Each subsection is carefully designed to address specific aspects of open data stewardship, providing a comprehensive educational framework for individuals keen on excelling in the dynamic and critical field of open data.

The following figure presents an overview of the sample curriculum presented in the following subsections:

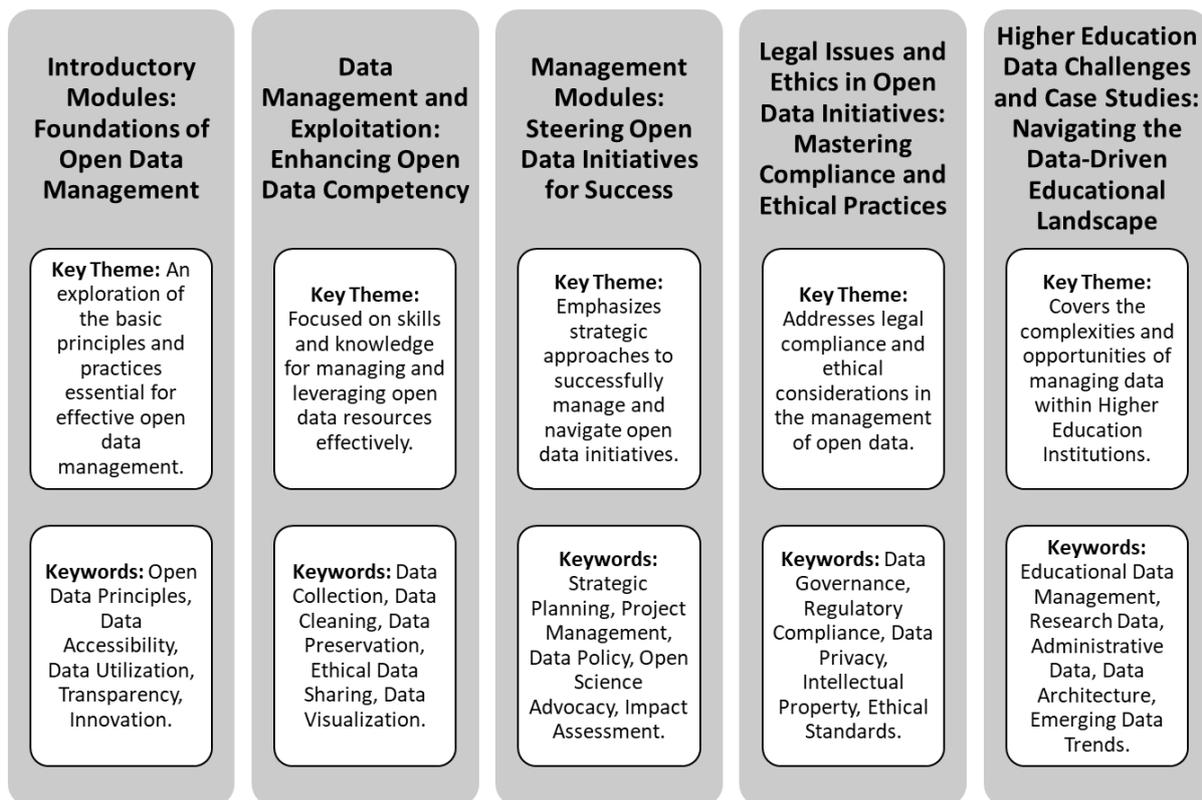

*Figure 2: Overview of sample curriculum for open data stewards.*

*Introductory Modules: Foundations of Open Data Management*

In the context of data accessibility and utilization, open data is recognized for its role in promoting transparency, empowerment, and innovation. It enables individuals to make informed decisions, encourages collaboration among diverse communities, and drives research and innovation across various disciplines. To fully realize the transformative potential of open data, we need to engage in a comprehensive exploration of the principles, policies, and practices essential for its effective management. The learning objectives of this knowledge stream are focused on to providing learners with a strong foundation in open data management principles, equip them with data science and programming skills, and introduce them to the fundamental aspects of efficient data management. By the completion of these modules, participants will have the ability to:

- Understand the core principles and policies that guide open data management, ensuring that our actions align with the values of openness, accessibility, and reusability.

- Master the fundamental concepts and techniques of data science and analytics, equipping one to transform raw data into actionable insights.

- Develop proficiency in programming languages relevant to open data analysis, enabling us to automate tasks and streamline workflows.

*Data Management and Exploitation: Enhancing Open Data Competency*

The knowledge stream on "Data Management and Exploitation" aims to equip learners with the skills and knowledge needed to effectively manage and leverage open data resources. This category comprises four essential modules, each designed to provide learners with a deep understanding of data collection and cleaning, data storage and preservation, data sharing and collaboration, as well as data analysis and visualization. Additionally, they will learn about data quality assessment and management, a critical aspect of open data projects.

The Learning Objectives of this knowledge stream are:

- Identify suitable data collection methods for specific research inquiries.
- Assess data quality to ensure its dependability and precision.
- Implement effective techniques for data cleaning and preprocessing.
- Gain insights into the ethical and legal aspects associated with data sharing.
- Choose appropriate data analysis methods based on data type and research objectives.
- Develop proficiency in data profiling, cleansing, and transformation to enhance data quality.
- Create compelling data visualizations to effectively communicate findings and insights to stakeholders.

*Management Modules: Steering Open Data Initiatives for Success*

The need for effective administration of open data initiatives arises from the rapid proliferation of data sources, diverse stakeholder interests, and evolving regulatory landscapes, demanding a strategic approach to ensure the success of these initiatives. The challenges surrounding data collection, quality assurance, storage, sharing, and analysis demands careful consideration and proactive management. Strategic management of data issues is crucial in this context, since it involves developing and implementing effective strategies that not only tackle current data concerns but also correspond with wider corporate objectives and social expectations. By recognizing the significance of steering open data initiatives through strategic management, organizations can unlock the full potential of their data resources, fostering transparency, innovation, and informed decision-making.

The Learning Objectives of this knowledge stream are:

- Develop a strategic mindset to identify, prioritize, and leverage open data opportunities aligned with organizational goals and objectives.

- Apply effective project management principles and practices to plan, execute, and monitor open data initiatives, ensuring operational excellence and impact.

- Understand and practice ethical data citation and attribution, recognizing the importance of proper credit and recognition in open data projects.

- Promote open science principles and advocate for data policies that uphold ethical and equitable open data practices.

- Master techniques for disseminating open data to diverse audiences, emphasizing data literacy and user engagement for wider reach.

- Measure the impact of open data initiatives and assess their success in achieving intended outcomes, fostering accountability and improvement.

*Legal Issues and Ethics in Open Data Initiatives: Mastering Compliance and Ethical Practices*

Understanding the legal and ethical implications of open data projects is extremely important for their appropriate utilization. This section explores the complex interplay between compliance with the law, ethical obligations, and effective handling of open data, highlighting the necessity of integrating open data strategies with both legal standards and ethical considerations, thereby ensuring judicious data management and building trust among diverse stakeholders. The modules in this part aim to provide learners with the essential knowledge and competences needed to proficiently navigate these intricate domains. Upon completing this knowledge stream, learners will be able to:

- Develop a comprehensive understanding of data governance, focusing on its legal and ethical dimensions in open data management.

- Identify and effectively mitigate potential risks associated with the management of open data, ensuring responsible handling.

- Apply ethical principles in practical data management scenarios, demonstrating a commitment to ethical standards in open data initiatives.
- Navigate the legal and regulatory landscape pertaining to open data, understanding its implications and ensuring compliance.
- Understand the critical role of open data standards and interoperability in promoting effective data sharing and usage across platforms and organizations.
- Foster a culture of transparency, accountability, and ethical conduct in the management and dissemination of open data.

*Higher Education Data Challenges and Case Studies: Navigating the Data-Driven Educational Landscape*

This section focuses on the diverse challenges and opportunities in managing data within HEIs, encompassing educational processes, research, administrative procedures, and the integration of emerging trends in open data management. The modules outlined below provide comprehensive insights into various aspects of data management in HEIs, equipping learners with the skills and knowledge to leverage data for strategic decision-making, compliance, and innovation in educational settings.

Upon completing this knowledge stream, learners will:

- Understand the complexities of managing educational data within HEIs, including data generated by Learning Management Systems (LMS) and other educational platforms.
- Gain insights into best practices for research data management across the research lifecycle, focusing on ethical, legal, and open science principles.
- Acquire knowledge of managing administrative data in HEIs, understanding its significance, and ensuring compliance with data governance standards.
- Explore the modern data infrastructure and architecture within HEIs, emphasizing open data principles and practices.
- Be aware about emerging trends in open data management, understanding their implications for HEIs and the broader educational sector.

**Conclusions**

Data plays a fundamental role in the paradigm of open science while open data acts as the cornerstone for scientific inquiry, innovation, and knowledge dissemination. By processing student records and academic transcripts, or their research data and administrative procedures, vast amounts of possible innovations, decision-making processes and transparency can be initiated. Therefore, in order to fully utilize this vast potential, it is necessary to have a dedicated team of open data stewards that has expertise in principles, techniques and practices related to open data management.

Open data stewards serve as the guardians of HEIs' data assets, ensuring that they are collected, managed, and shared responsibly. They are the catalysts of transformation, leveraging data to improve student outcomes, enhance faculty productivity, and optimize institutional efficiency. Furthermore, they act as advocates for openness, ensuring that data is accessible to all who can benefit from its insights, assuming eventually the role of knowledge custodians, ensuring that the data produced by HEIs is used ethically and responsibly, contributing to a more equitable and informed future.

This article aimed to explore and define the crucial skills and competences necessary for effective data stewardship in higher education, particularly focusing on the emerging role of open data stewards. The research was driven by the need to understand the evolving demands of open data and the implications for data management in academic settings.

Our research identified five key competence categories for open data stewards: Introductory Modules, Data Management, Strategic Management Modules, Legal Issues and Ethics, and Higher Education Data Challenges and Case Studies. These categories encompass a comprehensive framework addressing both theoretical knowledge and practical skills necessary for managing and promoting open data in universities.

This study contributes to the academic literature by providing a detailed framework for the training and development of data stewards in higher education. It emphasizes the importance of equipping these professionals with a broad range of competencies, aligning with the latest trends in open data practices and the evolving landscape of higher education.

The findings of this paper have significant implications for higher education institutions. By adopting the proposed curriculum framework, universities can enhance the skills of their data stewards, ensuring they are well-prepared to manage the growing complexity and volume of data in academic institutions. This advancement will facilitate improved data management practices, leading to increased transparency, collaboration, and innovation in research. The paper also highlights the transformative potential of open data stewards in shaping data accessibility and sharing practices in academia.

In conclusion, this article underscores the critical role of open data stewards in modern higher education and provides a roadmap for developing the necessary competencies to excel in this role. The proposed framework serves as a foundation for future curriculum development, aiming to address the demands of open science and foster a culture of effective data management in HEIs.

**Limitations and Future Work**

The limitations of this work are primarily twofold. Firstly, the findings and recommendations are grounded in a literature review, which, while comprehensive, may not capture the full scope of practical experiences and current practices in data stewardship within HEIs. Literature reviews synthesize existing information, but they can be limited by the data available in published works, which may not reflect the most recent developments or undocumented industry practices.

Secondly, the proposed curriculum and the role of Open Data Stewards, as outlined in this paper, have not been empirically tested through a pilot study. Without validation from actual implementation, there remains uncertainty regarding the curriculum's effectiveness in real-world settings. A pilot study would provide valuable insights into the practical challenges and outcomes of the curriculum when applied to the target audience, allowing for refinement and adjustments to improve its relevance and impact. The absence of such a study means that the proposed framework has yet to be proven in practice, and its efficacy is hypothesized rather than evidenced.

The future work for this project will concentrate on empirically validating the proposed curriculum for Open Data Stewards through pilot studies in various Higher Education Institutions (HEIs). These studies aim to test the curriculum's effectiveness, identify implementation challenges, assess its impact on participants' skills and competences, and gather feedback for curriculum refinement to better meet the needs of different HEIs.

Additionally, to enrich the theoretical foundation laid out by the literature review, future work should also focus on gathering and analyzing practical insights from current Open Data Stewards, data managers, and academic professionals. Including detailed case studies and interviews would provide a more comprehensive understanding of the practical aspects of data stewardship in HEIs and add valuable real-world perspectives to the research.

Moreover, engaging in collaborations with industry practitioners and academic experts in the field of open data and data management would provide an opportunity to validate and enrich the curriculum. These collaborations can help in aligning the curriculum with industry standards and expectations, as well as in identifying potential areas of improvement.

By focusing on these areas, future work can significantly enhance the quality and applicability of the research, ensuring that the curriculum for Open Data Stewards is both robust and relevant to the evolving needs of Higher Education Institutions.

**References**

Atenas, J., Havemann, L., & Priego, E. (2015). Open data as open educational resources: Towards transversal skills and global citizenship. Open praxis, 7(4), 377-389.

Atkins, D. E., Brown, J. S., & Hammond, A. L. (2007). A review of the open educational resources (OER) movement: Achievements, challenges, and new opportunities (Vol. 164). Mountain View: Creative common.

Borgerud, C., & Borglund, E. (2020). Open research data, an archival challenge? Archival Science, 20(3), 279-302.

Borgman, C. L. (2012). The conundrum of sharing research data. Journal of the American Society for Information Science and Technology, 63(6), 1059-1078.

Borgman, C. L. (2018). Open data, grey data, and stewardship: Universities at the privacy frontier. Berkeley Technology Law Journal, 33(2), 365-412.

Bunkar, A. R., & Bhatt, D. D. (2020). Perception of Researchers; Academicians of Parul University towards Research Data Management System; Role of Library: A Study. DESIDOC Journal of Library & Information Technology, 40(3), 139-146.

Cai, L., & Zhu, Y. (2015). The challenges of data quality and data quality assessment in the big data era. Data Science Journal, 14(0), 2. https://doi.org/10.5334/dsj-2015-002

Centre for Institutions and Governance (CIG). (2023). Centre for Institutions and Governance. Retrieved from https://www.bschool.cuhk.edu.hk/centres/centre-for-institutions-and-governance/

Coughlan, T. (2020). The use of open data as a material for learning. Education Technology Research & Development, 68, 383–411.

Curriki (2023). [Webpage]. Retrieved from https://curriki.org/

Donner, E. K. (2023). Research data management systems and the organization of universities and research institutes: A systematic literature review. Journal of Librarianship and Information Science, 55(2), 261-281.

European Commission. (2016). A European Open Science Cloud. Retrieved from https://research-and-innovation.ec.europa.eu/strategy/strategy-2020-2024/our-digital-future/open-science_en

European Commission. (2019). Guidelines on FAIR Data Management in Horizon 2020. Retrieved from https://ec.europa.eu/research/participants/data/ref/h2020/grants_manual/hi/oa_pilot/h2020-hi-oa-data-mgt_en.pdf

European Commission. (2021). European Open Science Cloud (EOSC). Retrieved from https://eosc.eu/

European Commission. (2023). Facts and figures for open research data. Retrieved from https://research-and-innovation.ec.europa.eu/strategy/strategy-2020-2024/our-digital-future/open-science/open-science-monitor/facts-and-figures-open-research-data_en

Fecher, B., & Friesike, S. (2014). Open science: one term, five schools of thought (pp. 17-47). Springer International Publishing.

Felden, J., Möller, L., Schindler, U., Huber, R., Schumacher, S., Koppe, R., ... & Glöckner, F. O. (2023). PANGAEA-Data Publisher for Earth & Environmental Science. Scientific Data, 10(1), 347.

Fitsilis, P., Iatrellis, O., & Tsoutsa, P. (2022, November). Using TOSCA language to model personalized educational content: Introducing eduTOSCA. In Proceedings of the 26th Pan-Hellenic Conference on Informatics (pp. 355-360).

Gallagher, R. V., Falster, D. S., Maitner, B. S., Salguero-Gómez, R., Vandvik, V., Pearse, W. D., ... & Enquist, B. J. (2020). Open Science principles for accelerating trait-based science across the Tree of Life. Nature ecology & evolution, 4(3), 294-303.

Gillenson, M. L. (2023). Fundamentals of database management systems. John Wiley & Sons.

Hansen, J. D., & Reich, J. (2015). Democratizing education? Examining access and usage patterns in massive open online courses. Science, 350(6265), 1245-1248.

Hansson, K., & Dahlgren, A. (2022). Open research data repositories: Practices, norms, and metadata for sharing images. Journal of the Association for Information Science and Technology, 73(2), 303-316.

Hendriyati, P., Agustin, F., Rahardja, U., & Ramadhan, T. (2022). Management information systems on integrated student and lecturer data. Aptisi Transactions on Management (ATM), 6(1), 1-9.

Hesteren, D., & Knippenberg, V. (2021). OPEN DATA MATURITY REPORT 2021. Luxembourg: Publications Office of the European Union.

Hylén, J. (n.d.). Open educational resources: Opportunities and challenges. OECD's Centre for Educational Research and Innovation. Retrieved from www.oecd.org/edu/ceri

Iatrellis, O., Kameas, A., & Fitsilis, P. (2017). Academic advising systems: A systematic literature review of empirical evidence. Education Sciences, 7(4), 90.

Iatrellis, O., Kameas, A., & Fitsilis, P. (2019). A novel integrated approach to the execution of personalized and self-evolving learning pathways. Education and Information Technologies, 24(1), 781-803.

Iatrellis, O., Savvas, I. K., Kameas, A., & Fitsilis, P. (2020). Integrated learning pathways in higher education: A framework enhanced with machine learning and semantics. Education and Information Technologies, 25, 3109-3129.

Ismael, S. N., Mohd, O., & Abd Rahim, Y. (2018). Implementation of open data in higher education: A review. Journal of Engineering Science and Technology, 13(11), 3489-3499.

Janssen, M., Charalabidis, Y., & Zuiderwijk, A. (2012). Benefits, adoption barriers and myths of open data and open government. Information systems management, 29(4), 258-268.

Karmanovskiy, N., Mouromtsev, D., Navrotskiy, M., Pavlov, D., & Radchenko, I. (2016). A case study of open science concept: linked open data in university. In Digital Transformation and Global Society: First International Conference, DTGS 2016, St. Petersburg, Russia, June 22-24, 2016, Revised Selected Papers 1 (pp. 400-403). Springer International Publishing.

Kassen, M. (2018). Adopting and managing open data: Stakeholder perspectives, challenges and policy recommendations. Aslib Journal of Information Management, 70(5), 518-537.

Khan Academy (2023). [Webpage]. Retrieved from https://www.khanacademy.org/

Krawitz, M., Law, J., & Litman, S. (n.d.). Many college and university leaders remain unsure of how to incorporate analytics into their operations: What really works? McKinsey & Company. Retrieved from https://www.mckinsey.com/industries/education/our-insights/how-higher-education-institutions-can-transform-themselves-using-advanced-analytics

Leong, K., Sung, A., Au, D., & Blanchard, C. (2020). A review of the trend of microlearning. Journal of Work-Applied Management, 13(1), 88-102.

Maharana, K., Mondal, S., & Nemade, B. (2022). A review: Data pre-processing and data augmentation techniques. Global Transitions Proceedings, 3(1), 91-99.

McDonald, J. (2022). Universities, rich in data, struggle to capture its value, study finds. UCLA Newsroom. Retrieved from https://newsroom.ucla.edu/releases/universities-struggle-to-leverage-data

Merlot (2023). [Webpage]. Retrieved from http://www.merlot.org/

Mons, B. (2018). Data stewardship for open science: Implementing FAIR principles. CRC Press.

Nag, M. B., & Ahmad Malik, F. (2023). Data analysis and interpretation. In Repatriation Management and Competency Transfer in a Culturally Dynamic World (pp. 93-140). Singapore: Springer Nature Singapore.

National Academies of Sciences, Engineering, and Medicine. (2018). Open Science by Design: Realizing a Vision for 21st Century Research. Washington, DC: The National Academies Press.

OER Commons (2023). [Webpage]. Retrieved from https://oercommons.org/

Open Knowledge Foundation. (2017). Open Definition 2.1-Open Definition-Defining Open in Open Data, Open Content and Open Knowledge.

OpenCourseWare Consortium (2023). [Webpage]. Retrieved from https://ocw.mit.edu/

Pampel, H., & Dallmeier-Tiessen, S. (2014). Open research data: From vision to practice. Opening science: The evolving guide on how the Internet is changing research, collaboration and scholarly publishing, 213-224.

Papageorgiou, G., Loukis, E., Pappas, G., Rizun, N., Saxena, S., Charalabidis, Y., & Alexopoulos, C. (2023, September). Open Government Data in Educational Programs Curriculum: Current State and Prospects. In International Conference on Business Informatics Research (pp. 311-326). Cham: Springer Nature Switzerland.

Pascu, C., & Burgelman, J. C. (2022). Open data: The building block of 21st century (open) science. Data & Policy, 4, e15.

Perkmann, M., & Schildt, H. (2015). Open data partnerships between firms and universities: The role of boundary organizations. Research Policy, 44(5), 1133-1143.

Piedra, N., Tovar, E., Colomo-Palacios, R., Lopez-Vargas, J., & Alexandra Chicaiza, J. (2014). Consuming and producing linked open data: the case of Opencourseware. Program, 48(1), 16-40.

Radchenko, I., Koroleva, A., & Baranov, Y. (2018). On Portrait of a Specialist in Open Data. arXiv preprint arXiv:1805.07598.

Ramírez, Y., & Tejada, Á. (2018). Corporate governance of universities: improving transparency and accountability. International Journal of Disclosure and Governance, 15, 29-39.

Reichman, O. J., Jones, M. B., & Schildhauer, M. P. (2011). Challenges and opportunities of open data in ecology. Science, 331(6018), 703-705.

Ridzuan, F., & Zainon, W. M. N. W. (2019). A review on data cleansing methods for big data. Procedia Computer Science, 161, 731-738.

Rodriguez-F, I. E., Arcos-Medina, G., Pástor, D., Oñate, A., & Gómez, O. S. (2020, October). Open Data in Higher Education-A Systematic Literature Review. In The International Conference on Advances in Emerging Trends and Technologies (pp. 75-88). Cham: Springer International Publishing.

Rodriguez-F, I. E., Arcos-Medina, G., Pástor, D., Oñate, A., & Gómez, O. S. (2020). Open Data in Higher Education-A Systematic Literature Review. In Advances in Emerging Trends and Technologies: Proceedings of ICAETT 2020 (pp. 75-88).

Rosenbaum, S. (2010). Data governance and stewardship: designing data stewardship entities and advancing data access. Health services research, 45(5p2), 1442-1455.

Shu, X., & Ye, Y. (2023). Knowledge Discovery: Methods from data mining and machine learning. Social Science Research, 110, 102817.

Silveira, I. F. (2016). OER and MOOC: The need for openness. Issues in Informing Science and Information Technology, 13, 209-223.

Times Higher Education. (2023). World University Rankings 2024 methodology. Retrieved from https://www.timeshighereducation.com/world-university-rankings/world-university-rankings-2024-methodology

Tran, E., & Scholtes, G. (2015). Open data literature review. Berkeley School of Law, University of California.

Tzitzikas, Y., Pitikakis, M., Giakoumis, G., Varouha, K., & Karkanaki, E. (2021). How can a university take its first steps in open data?. In Metadata and Semantic Research: 14th International Conference, MTSR 2020, Madrid, Spain, December 2–4, 2020, Revised Selected Papers 14 (pp. 155-167). Springer International Publishing.

U.S. Department of Education, National Center for Education Statistics. (2023), Integrated Postsecondary Education Data System (IPEDS). Retrieved from https://nces.ed.gov/ipeds/

U.S. Department of Education. (2023). College Navigator [Web application]. Retrieved from https://nces.ed.gov/collegenavigator/

UNESCO (2023, December 21). Open Educational Resources (OER): Benefits, Recommendation, and UNESCO's Promotion. Retrieved from https://www.unesco.org/en/open-educational-resources

UNESCO. (n.d.). Understanding Open Science. [Open science definition]. UNESCO Recommendation. Retrieved from https://unesdoc.unesco.org/ark:/48223/pf0000383323

Universities UK, & Open Data Institute. (2015). Open data in higher education: An introductory guide. Retrieved from https://dera.ioe.ac.uk//id/eprint/26183

University of California. (2023). Data sharing Policies and Tools. Retrieved from https://osc.universityofcalifornia.edu/for-authors/open-data/

University of Cambridge. (2023). Data management and sharing. Retrieved from https://osc.cam.ac.uk/open-research/data-management-sharing

Vicente-Saez, R., Gustafsson, R., & Van den Brande, L. (2020). The dawn of an open exploration era: Emergent principles and practices of open science and innovation of university research teams in a digital world. Technological Forecasting and Social Change, 156, 120037.

West, J. D., & Paton, G. (2018). Learning analytics for open and distance education. Journal of Learning Analytics, 5(1), 1-6.

Wilkinson, M. D., Dumontier, M., Aalbersberg, I. J., Appleton, G., Axton, M., Baak, A., ... & Mons, B. (2016). The FAIR Guiding Principles for scientific data management and stewardship. Scientific Data, 3(1), 160018. https://doi.org/10.1038/sdata.2016.18

Zubcoff, J., Vaquer Gregori, L., Mazón, J. N., Maciá Pérez, F., Garrigós, I., Fuster-Guilló, A., & Cárcel Alcover, J. V. (2016). The university as an open data ecosystem.

Zuiderwijk, A., Shinde, R., & Jeng, W. (2020). What drives and inhibits researchers to share and use open research data? A systematic literature review to analyze factors influencing open research data adoption. PloS one, 15(9), e0239283.

Zhu, Y.J., Freund, L. (2020). Exploring Open Data Initiatives in Higher Education. In: Sundqvist, A., Berget, G., Nolin, J., Skjerdingstad, K. (eds) Sustainable Digital Communities. iConference 2020. Lecture Notes in Computer Science, vol 12051. Springer, Cham. https://doi.org/10.1007/978-3-030-43687-2_60